# Serious Games in Digital Gaming: A Comprehensive Review of Applications, Game Engines and Advancements

ALEXANDROS GAZIS[1*], ELEFTHERIA KATSIRI[1,2]
[1]Democritus University of Thrace, School of Engineering,
Department of Electrical and Computer,
Engineering, Xanthi, 67100,
GREECE
[2]Institute for the Management of Information Systems,
Athena Research & Innovation Center in Information Communication & Knowledge Technologies,
Marousi, 15125,
GREECE

*Corresponding Author*

*Abstract:* - Serious games are defined as applied games that focus on the gamification of an experience (e.g., learning and training activities) and are not strictly for entertainment purposes. In recent years, serious games have become increasingly popular due to their ability to simultaneously educate and entertain users. In this review, we provide a comprehensive overview of the different types of digital games and expand on the serious games genre while focusing on its various applications. Furthermore, we present the most widely used game engines used in the game development industry and extend the Unity game machine advantages. Lastly, we conclude our research with a detailed comparison of the two most popular choices (Unreal and Unity engines) and their respective advantages and disadvantages while providing future suggestions for serious digital game development.

## 1 Introduction

Digital games can bring our dreams to life and this distinctive feature separates them from other forms of interactive entertainment. Specifically, their notable difference from other sources of information/education/entertainment, such as books and films, is that their potential is tightly connected to the imagination of the individual. This happens as they do not allow the viewer to engage and interact thus feeling he is an active member of an online virtual community, [1], [2].

In recent years, digital games have become an integral part of our everyday life and due to the increase of smartphones and smart devices, the new generation is spending more and more time on so-called "gaming". For this reason, it is necessary to further study and thoroughly understand the characteristics of the digital games that young people like to create games with similar qualities. The aim of researchers should be to understand the behavior, opinions, and political beliefs of the younger generations and focus on the psychological effects of games, [3] not only emphasizing violent games, [4]. According to the current literature, it has been shown that digital games can have a positive impact on both mood and behavior of players, [5], [6], [7].

Furthermore, "serious games" (SG) are a genre that focuses on story-telling experience "outside the context of entertainment, where the narration progresses as a sequence of patterns impressive in quality ... and is part of a thoughtful progress", [8] i.e., applications that aim to "gamify" learning processes in a modern learning environment, [9]. In particular, this is achieved by creating an application that simulates a -usually learning- process or through the construction of virtual worlds and communities alike. These games aim to promote interactive and experiential learning so that players, in their efforts to win the game, gain a rudimentary knowledge of various concepts and experiences. These games can significantly help in the development of critical thinking and the ability to solve problems rapidly as, through feedback (usually through a point or life collection system), they reward or not each player's choices and thus educate them on correct decisions. Therefore, in the context of exploring a virtual world, the player is educated, learns, chooses, and decides what is a correct decision in a realistic (real-life) manner.

In addition, there are 2 categories of players based on their gameplay style, the "core" and the "casual", [10]. More specifically, the first type of player refers to people who usually spend more hours playing a game and enjoy beating each quest/level. This category enjoys the increased difficulty in a game and can overcome multiple obstacles and challenges in each game to move to the next level. In contrast, the second type prefers easier games and enjoys the overall experience of the game with medium to low-difficulty gameplay.

Our study focuses on providing a comprehensive look at the current state of serious games in digital gaming, with insights into their applications, game engines, and advancements, as well as future directions for the industry. In the next sections, firstly we provide a literature review of digital gaming. Then, we define the different genres of games and compare digital games with serious games providing real-life examples of released games. Secondly, we present the most widely used game engines used. Thirdly, we focus on the advantages of the Unity engine and its middleware architecture, arguably one of the most commonly used engines for game development worldwide. In addition, we compare Unity with Unreal, i.e., the two most widely used game engines. Lastly, we conclude our study and suggest future research steps based on our publication.

## 2 Related Work

Worldwide, the creation and adoption of educational computer games have been rapidly and steadily increasing. In particular, the introduction of the use of electronic devices in everyday life and the increase in their capabilities, combined with their decreasing price over the years (Moore's Law), [11] has provided a large part of the population with accessibility to digital consoles able to play complex and highly demanding (in computer resources) games. Specifically, in our everyday life, a cheap electronic device (tablet or mobile) can support a multitude of different types and categories of games, offering the owner the possibility to enhance their game experience with low-cost devices.

In particular, digital games are a highly successful and popular means of engaging the interest of young people - a particularly difficult task to achieve through traditional practices - and are often used as a means of studying how they interact with society, [12]. More specifically, they can contribute to shaping young people's behavior and assist in learning about the principles and functioning of societies. As mentioned in [13], digital games can encourage young people to take on new roles as well as social identities (profiles) by providing strong incentives to receive and understand new information and knowledge in general. It is worth mentioning that the gamer's age is particularly susceptible to receiving new communicative stimuli as well as learning from new experiences, [14]. Finally, these options may result in changing their respective behavior, their way of thinking, their opinions, and in some cases even their political beliefs.

Moreover, one of the key issues that must be taken under consideration when designing a game is the characteristics of the audience (age, gender, culture) and the main types of players it will address. In addition, concerning the technical characteristics of a game, it is particularly important to consider the technological tools that will be used, such as the development platform (Windows, Xbox, Mac, PlayStation), the game engine, the navigation experience in the game, the elements (graphics, sound), the way and the number of options that the player can enter and the overall interaction with the game. In this respect, it is highlighted that serious games are a means of education for both minors and adults depending on the design and subject matter of each game.

### 2.1 Categories of Digital Games

Different genres of games exist based on both their content and the way each player interacts with the game, but typically digital games can be categorized into one of the following [15]:

1. <u>Strategy games</u>: involve the careful planning and efficient resource management of small virtual worlds/online communities. Specifically, this type usually aims at mature (age) audiences and is usually referred to as "thinking games".
2. <u>Real Role-playing Games</u>: this category has similar characteristics to "Strategy games". The notable difference in this category is that the user can use multiple characters with different skills, emphasizing the development of a character (player avatar/player profile) through a reward system. Specifically, the player aims to upgrade their character through the accumulation of points and promotion to different "classes".

3. <u>Massively Multi Role Playing Games</u>: they consist of multiple virtual worlds and involve the parallel interaction of multiple users. These game worlds are highly interesting for researchers who aim to study user behavior. Specifically, due to the emergence of Web 2.0, [16] and the subsequent social networking phenomenon, the creation of a Multi-User Virtual Environment (MUVE) has become particularly simple as it depends solely on 3 distinct processes: hosting server, software/interfaces for user communication with the game and user identification. Upon identification and successful login, the user is part of the digital world and can communicate and exchange information with other members of the small virtual worlds/online communities.
4. <u>Simulation games</u>: they aim to simulate a specific activity realistically, taking into account not only the laws of physics but also the real-life constraints and operating rules of everyday life (i.e., the real world).
5. <u>Government Simulation Games</u>: while this category of games has identical characteristics to both "Simulation Games" and "Strategy Games", due to their unique features, i.e., discouraging the player from making arbitrary decisions and removing overall elements of free action and navigation, they are considered a separate game genre.

## 2.2 Digital Games and Serious Digital Games
Digital games are categorized into various categories depending on their purpose of creation, [17]:
1. Commercial games designed to entertain and amuse the user
2. Serious games designed for learning purposes.

Similarly, they are divided into the following categories, depending on their design focus, [18]:
1. Design is driven by a specific didactic approach,
2. Design without an educational purpose but with the possibility for users to learn.
3. Design based on a specific learning theory but without a specific teaching approach.

The third category can be used for educational purposes without the intervention of a teacher. Indicatively, some examples of digital games, [19] are "Sims Pets" which involves the development of problem-solving skills, "Age of Empires III" for the development of social skills and "Harry Potter and the Goblet of Fire" for the development of writing skills.

According to [20], [21], SGs are defined as digital games aiming to be used as learning models rather than as a means of user entertainment. In particular, SGs have been widely used in various fields of human activity to train users. In this study, the focus is on SGs, i.e. games that were developed and used exclusively for educational purposes. In particular, an SG in the field of education is defined as an application in which: "the player competes cognitively with a computer system according to certain predefined rules. These games aim to use the user's entertainment as a means of learning concepts related to government or business, education, health, public policy, and strategic communication objectives", [22]. Therefore, the purpose of SGs is to motivate players in the light of an educational environment to transmit ideas, and values, and often to motivate them to a specific action or to think about a specific concept/idea, [23]. In this way, the players' behavior is changed as they think about their actions and their possible impact on real-life situations in everyday life i.e., the real world, [24]. SGs have been widely used in the last decade, [25] in the field of education as they offer the following, [26]:
1. Enhancing the instructor-trainee interaction
2. Increase the instructor's concentration, [27]
3. Improve critical thinking and logical thinking.

## 2.3 Serious Digital Games Applications
SGs help students to apply factual knowledge, learn on demand, adapt to new situations, and, gain new experiences and knowledge, [28]. SGs assist in making the learning process more effective and efficient as they maintain high levels of interest among participants and promote the active participation of users in problem-solving. As a result, SGs are a particularly important training tool as they follow the so-called situated learning model, [29], [30], [31]. Over the years, SGs continuously incorporate new features and are enriched with new game and graphical methods.

Finally, some indicative applications of SGs include corporate training, [32], entrepreneurship, [33], foreign, [34] and programming, [35], [36] language learning, cultural heritage, [37], ecological awareness, [38], [39], military service, [40], [41], pharmacology [42], health, [43], [44], [45] and health care, [46], circular economy, [47], stress management, [48], engineering risk management [49], cyber security, [50] and law enforcement, [51], [52], [53].

# 3 Game Engines for Game Development

Many different game engines exist that are used depending on the needs of the project. In this section, we will present the most widely used game engines and focus on Unity.

## 3.1 Defining: Game Engines as Middleware Software Solutions

There are 2 ways to develop a game, [54], [55]:
   a. the traditional way of development where the developer develops the game from scratch using a programming language or
   b. the modern approach which is based on software packages, i.e., a framework of software development tools that deal with predefined game functions.

More specifically, in the last decade, many application packages, also known as frameworks, have emerged which have provided the software engineer (developer) with useful tools and methods to assist at all stages of the game development, [56], [57]. These tools, in addition to assisting in the design and implementation process, provided a way of modeling and standardizing many continuous integration and continuous operation processes. As a result, the application frameworks contributed in the long run to reduce the necessary time for development, maintenance, and deployment time of the applications, [58], [59], [60].

These middleware approaches, in the field of digital game development and design, are the so-called game engines that provide the necessary tools to create a game regarding its basic functions such as:
- the player's movement,
- the sound,
- the development and code execution interfaces, etc.

In particular, these middleware provide a bundle for all the necessary software packages as well as all the predefined rules and methods regarding the following:
- networking of player computers,
- operation of the graphics of an application,
- or, more generally, various functions such as the development platform of the digital game (windows/Linux).

More specifically, these software properties, although they are purely related to the early stages of game development, i.e., the so-called backbone-core of the application, they are of paramount importance for the development of digital games with continuous integration and continuous delivery of the information flow within the schedule, [61]. With the use of machines, developers focus their development efforts and time purely on the application itself and do not have to "reinvent the wheel" i.e., deal with questions such as what communication protocol the application will work, what are the possible login/verification options for the user, what will be the resolution of the game depending on the execution surface (screen), etc.

In addition, a particularly interesting feature of game engines as middleware entities is the "game editor", i.e., the graphical tool that allows the creation of a game usually through a drag-and-drop menu with multiple elements and different options. Unfortunately, this feature has several limitations but, in the early stages of a game's development, it allows developers to easily and quickly create initial functional prototypes. Finally, this is particularly important as it allows non-technical members of the game development team, such as graphic designers and other specialists, to start contributing to the game development project from the early stages of the project. In this way, non-technical members' working hours are not wasted waiting for long periods, i.e., until the developers have completed the programming and the necessary testing-testing of the prototype versions of the programming code for the execution of the game.

## 3.2 Defining: Game Engines Frameworks

Specifically, there are several digital game development engines, of which the most widely used are the following:
   1. OpenSimulator, [62] an open-source software engine with capabilities of running on multiple platforms, multi-user interfacing, and running in 3D virtual world environments. In particular, this engine can be used to create virtual environments in which users can access a server through the use of clients of different interface protocols. In particular, it has a very useful optional installation (Hypergrid) that allows users to visit other active OpenSimulator installations on the Internet. This is accomplished through the software license installed on a personal computer, which creates a kind of distributed network of computers (Metaverse type). OpenSimulator is written in C# and runs both on Windows via the .NET framework and on Unix software machines via the Mono framework. The source code is made available for free through a Berkeley Software Distribution (BSD) license. In particular, this code may be used as-is or incorporated into existing commercial products without any restriction on use or future requirement, thus enhancing the integration of OpenSimulator into commercial applications and products.

2. Unreal Engine, [63] one of the best-known game engines developed by Epic Games and first used in the late 1990s in the "first-person" game Unreal. Subsequently, this engine was further developed and focused on games that mainly involved shooters or "first-person" mode games. Its ease of use as well as the multitude of tools it provided quickly allowed it to be used in "Massively Multi-Role Games" as well as "Real Role-Playing Games". The Unreal engine is written in the C++ programming language, it is open source, and due to the language chosen for its development, it is characterized by high speed and a high degree of portability. This means that it provides the possibility of migration to different environments and platforms without the occurrence of compatibility problems. This engine operates under a subscription service (monthly payment) and supports several elements such as fully dynamic lighting modes, instant game updates without interrupting game execution, full-screen game viewing during software development, interactive visualization tools for code flow, and sophisticated bug detection and debugging methods.

3. CryEngine, [64], [65] a game engine technology developed by Crytek. It was originally developed as part of Ubisoft's Far Cry game which was released in 2004 and was a game landmark in terms of the realism of graphics as well as the detail of objects in small virtual worlds/online communities. The latest version of CryEngine supports multiple elements such as natural lighting manipulation, dynamic/painted shadows, dynamic global illumination of game world objects in real-time, automated grid generation for navigation and user exploration via artificial intelligence systems, automated motion blur, depth of field switching and stereoscopic 3D support for all platforms. CryEngine engine technology has been used in recent years to create games on various software platforms, including Windows, Linux, OS X, Xbox, Play Station, Wii, iOS, and Android as it is particularly popular in 3D game development.

4. Unity, [66] composes a game engine capable of running on multiple platforms and was developed by Unity Technologies in 2005. In particular, it is characterized by a multitude of features such as Mecanim, an animation system for representing any character or object, real-time shading and lighting options for all development platforms, system updates with particle collision control to avoid bugs, and dynamic disable-enable links between objects and obstacles. Additionally, it is worth pointing out that while the Unity game engine is not as powerful as CryEngine or Unreal, it stands out due to its efficiency as well as its multitude of settings for publishing digital games across multiple platforms, [67], [68]. In particular, the aforementioned elements are worth noting as by using Unity, developers select and focus their efforts on developing code on a specific platform without spending hours configuring implementations to make the application run on other platforms. This capability makes software development cycles less time-consuming and rapidly agile thus ultimately increasing development efficiency throughout the project lifecycle. In addition, as of 2018, the engine has been extended to be able to support 27 platforms including iOS, Android, Windows, BlackBerry 10, OS X, Linux, Internet browsers, Flash, PlayStation 3, PlayStation Vita, Xbox 360, Windows Phone 8 and Wii U. Finally, Unity can be used to create 3D and 2D games as well as simulations for the majority of commercial products and platforms available on the market, [69].

Finally, it is emphasized that there is a particularly large number of available digital game development engines that are extensively used in the industry such as Game Maker, [70], [71] or the more recent Godot, [72], [73] but, in the context of this technical report we chose to briefly mention the most widely used in the industry.

## 3.3 Defining: Unity Game Engine

Unity is used worldwide in both industrial productions and academic research projects. One of the main reasons for its choice is that it provides a full version freely without any limitation or fee, as long as the annual earnings of the game project do not sum up to more than 100,000 US dollars, [74]. This amount is difficult to reach as during the early prototype versions of the games, developers may harness the full power of the engine but, it is difficult to scale their application that rapidly and have annual earnings of that amount. At the same time, unlike its competitor (Unreal), although it requires more computational resources for the same processes, it provides the possibility of easier export of the final exported file to a multitude of different computer systems without requiring multiple simultaneous developments on different platforms, [75], [76], [77].

When installing the Unity engine to a workstation, a developer will be forced to work with the following software components:
1. Unity Editor: the tool used for developing game programming objects, connecting game graphics/audio, and overall, the basic functionality and processes of our game.
2. Unity Hub: an application that provides the ability to manage Unity Projects, and install and monitor all the necessary elements of our game. As an example, it is worth mentioning that this management application provides a multitude of settings such as the configuration of the Unity version, support for different exported files (builds) on different platforms, the version of Visual Studio, and the IDE for development and debugging.

Furthermore, Unity handles its game aspect as a separate object or game component, [78]. Analytically, all object properties, i.e., the behaviors in a virtual world of a digital game, are determined by automated programs (scripts) that assign properties to objects such as movement and rotation. Specifically, in the Unity engine these programs have been developed using the C# language, which is a general-purpose object-oriented language created and maintained by Microsoft, [79]. More specifically, in the Unity engine, the properties of the architectural design principles of the middleware used consider the individual attributes of game objects to be stored as follows, [80]:
- Project: refers to the specific folder all the elements related to the digital game as well as its components are stored (libraries, files, etc.)
- Assets: refers to the actual files and sub-folders of a software project such as image files, audio files, programs, etc.
- Objects: are entities with which the player interacts. Each object in a game may contain information regarding its behavior such as lighting options, programs to be executed, etc.
- Components: i.e., the properties of an object that can be accordingly modified to give the necessary structure and function to the digital game. These are the cornerstone pieces that are used in a game and combine all the above options to determine the final behavior of the game.

Finally, it is worth mentioning that the Unity engine supports event-driven development. In particular, this software method is based on the operating principle that after the game is started, programs can run when a certain event (defined by the developer) occurs and add new functions and features. More specifically, the above element is particularly important as it makes the digital game being developed less demanding in terms of computational resources. It also adds interactive elements to the player as well as more immediacy and interactivity since his choices determine the final actions of the game. As a result, the game is not a monolithic application with predefined actions but a dynamic application that adapts to every playing style and players' personal decisions, [81].

Lastly, it is worth noting that Unity's Projects are usually structured via convention over configuration and the do not repeat your architectural approaches. This means that the structure of actual development files is usually as follows (separation of concerns):
- Assets: storage space for the components used during the game process that relate to all the functions of the core quests performed by the user.
- Prefabs: where the predefined game objects are stored. These objects are used multiple times during the gameplay and are replicated in multiple parts of our game (e.g., graphics for the missions, player interaction effects, etc.).
- Scenes: the actual scenes of our game, i.e., the various menus and levels of our game.

- Scripts: the programming codes, i.e., the so-called C# programs (scripts .cs) of our game.
- All other folders are not used for more than 1 object, so they are stored individually.

### 3.4 Defining: Unity vs Unreal Game Engine
Regarding the game engines that exist, we studied several of the aforementioned game engines, [82], [83], [84], [85], [85], [86] and concluded that the best choices in terms of performance, features, and software cost were Unity and Unreal, [87], [88]. In particular, after studying the software architecture as well as the capabilities of each solution, the Unity engine was chosen for the following reasons:

- It provides a free development license for gross annual developer/company revenues of less than 100,000 US dollars (USD), [74]
- It offers easy migration and porting of the application to different platforms
- It has been developed in a programming language supported by the well-known software company Microsoft therefore the engine has excellent documentation and is regularly updated.
- The use of C# (while Unreal in C++) provides the possibility of easier extension of the game, especially in case of making it available to the public via cloud infrastructure as, by using C# one can support both the back-end and the front-end of the application. More specifically, through C# developers can create an infrastructure in .NET, [89] using Azure SQL Server (back-end), [90] that will host the Unity digital game (application) and provide a feedback menu (communication screen) with the player on a website using Asp.Net or other well-known frameworks such as Blazor (front-end), [91]. That is, the Microsoft software and application ecosystem can cover all aspects of the development, deployment, and upgrading of this application.

In addition, for the selection of the Unity engine, emphasis is placed on the following features, [92], [93]:
- Free license for the development of digital games which is only slightly inferior in terms of features to the professional-commercial distribution of the engine
- Support for the development of any type of game from two to three dimensions in all available game categories
- Existence of an extensive software ecosystem through the provision of several features for setting lighting, graphics, sound, etc.
- Provide editing tools for scenarios, images, and audio and optimize the performance of two-dimensional models
- Providing a standardized asset pipeline, i.e., game properties such as application code and audiovisual effects can be compressed and reused across different platforms and environment versions. This is important as it allows easier development and maintenance of legacy versions (dependency tracking)
- Participation in a highly active online community of developers, [94].
- Machine support for LTS releases and more general software updates regularly, which are usually accompanied by major fixes to software security vulnerabilities
- Encourage the adoption of modern agile software development approaches through the iterative development model
- Ability to create a game on any platform: Mac, Windows, WebGL, Android, etc. Special mention should be made of the support of the WebGL platform, a software interface (API) in JavaScript that allows the execution of a multitude of games through simple web browsers, [95]. Furthermore, it is stressed that although the -actual- size of digital game files increases significantly when they are converted (built) to this platform, the possibility of fully integrating games into web browsers, at a time when mobile devices have dominated the market, is probably the future of many digital games
- Support the development of new forms of digital games using innovative technologies such as virtual reality (e.g., using special glasses)

## 4 Conclusion
Our study provided a comprehensive overview of what is a digital game, its different genres, and the importance of serious games in digital gaming. Furthermore, while focusing on serious games, we presented the most popular game engines and provided their key features as new technological advancements in the industry. Moreover, we have made a detailed comparison of the two most widely used game engines, Unreal and Unity, and highlighted the pros and cons of each approach while also discussing the advantages of the Unity engine and its middleware architecture. Lastly, throughout this article, we have presented real-life examples of released games and methods to illustrate the different genres of games and compare digital games and serious games alike.

Overall, our review emphasizes the potential of serious games to educate and entertain users, as well as future directions for the industry. This study can be a valuable resource for game developers, educators, and researchers interested in serious games and their potential applications.

## 5 Future Research

We suggest that future research should focus on exploring the effectiveness of serious games in various fields and the impact of emerging technologies, such as virtual reality and cloud gaming. According to the current literature, increasing use of Virtual Reality technology in existing infrastructures has been observed. We believe that due to the extensive software support and the large number and active participation of the Unity community, this engine will most likely support -without backward compatibility issues- the transition of existing games to cloud-based infrastructures. Cloud infrastructures are important as they will shift the game workload from a local computer to a remote server thus providing the opportunity to game to a bigger audience. This is extremely important as it will enable SGs to reach people with limited access to funding that can only use low-power and low-cost computers to educate. As a result, this feature will enhance the main goal of serious games, namely, the gamification of the learning experience, and will pave the way for focusing on new technologies in SGs such as Virtual Reality, [96], [97], [98].

**Contribution of Individual Authors to the Creation of a Scientific Article (Ghostwriting Policy)**
- Alexandros Gazis, was responsible for conceptualization, investigation, methodology, software, validation, visualization, writing the original draft, review-editing resources, carrying out simulations and writing the original draft.
- Eleftheria Katsiri, contributed to the conceptualization, formal analysis, funding acquisition, investigation, methodology, project administration, resources, supervision, validation, visualization, review, and editing of the original draft.